\documentclass[conference]{IEEEtran}
\usepackage[left=0.75in,right=0.75in,top=0.75in,bottom=0.83in]{geometry}
\usepackage[T1]{fontenc}
\usepackage[latin9]{inputenc}
\usepackage{amsthm}
\usepackage{amsmath} 
\usepackage{graphicx} 
\usepackage{color} 
\usepackage{cite}
\usepackage{algpseudocode}
\usepackage{algorithm}
\usepackage{amssymb}
\usepackage{subfigure}
\usepackage{esint}
\usepackage{algpseudocode}
\usepackage{epstopdf}
\usepackage{multirow}
\usepackage{makecell}
\usepackage{float}
\usepackage{lipsum}
\usepackage{balance}

\newtheorem{definition}{Definition}
\newtheorem{remark}{Remark}

\newtheorem{example}{Example}

\theoremstyle{plain}
\theoremstyle{plain}
\newtheorem{theorem}{Theorem}

\newcommand{\comment}[1]{}

\IEEEoverridecommandlockouts

\begin{document}

\title{\vspace{+0.25in}Asymmetric LOCO Codes: Constrained Codes \\ for Flash Memories}

\author{
   \IEEEauthorblockN{Ahmed Hareedy and Robert Calderbank}
   \IEEEauthorblockA{Electrical and Computer Engineering Department, Duke University, Durham, NC 27705 USA \\ ahmed.hareedy@duke.edu and robert.calderbank@duke.edu\vspace{-1.0em}}
}
\maketitle

%%%%%%%%%%%%%%%%%%%%%%%%%%%%%%%%%%%%%
\begin{abstract}
In data storage and data transmission, certain patterns are more likely to be subject to error when written (transmitted) onto the media. In magnetic recording systems with binary data and bipolar non-return-to-zero signaling, patterns that have insufficient separation between consecutive transitions exacerbate inter-symbol interference. Constrained codes are used to eliminate such error-prone patterns. A recent example is a new family of capacity-achieving constrained codes, named lexicographically-ordered constrained codes (LOCO codes). LOCO codes are symmetric, that is, the set of forbidden patterns is closed under taking pattern complements. LOCO codes are suboptimal in terms of rate when used in Flash devices where block erasure is employed since the complement of an error-prone pattern is not detrimental in these devices. This paper introduces asymmetric LOCO codes (A-LOCO codes), which are lexicographically-ordered constrained codes that forbid only those patterns that are detrimental for Flash performance. A-LOCO codes are also capacity-achieving, and at finite-lengths, they offer higher rates than the available state-of-the-art constrained codes designed for the same goal. The mapping-demapping between the index and the codeword in A-LOCO codes allows low-complexity encoding and decoding algorithms that are simpler than their LOCO counterparts.
\end{abstract}

%%%%%%%%%%%%%%%%%%%%%%%%%%%%%%%%%%%%%
\section{Introduction}\label{sec_intro}

Constrained codes are widely applied in various data storage systems to improve their performance. These codes were first employed in magnetic recording (MR) systems. Run-length-limited (RLL) codes \cite {tang_bahl} were used to extend the lifetime of peak detection in early IBM disk drives \cite{siegel_mr}. Binary RLL codes are typically used with bipolar non-return-to-zero inverted (NRZI) signaling, where a $0$ is represented by no transition, while a $1$ is represented by a transition. RLL codes are used to control the separation between consecutive transitions. Small separation exacerbates inter-symbol interference (ISI) while large separation results in losing synchronization at the receiver. Constrained codes also find application in modern MR systems \cite{vasic_prc} to improve sequence detection \cite{immink_1}.

We define the set $\mathcal{S}_x \triangleq \{010, 101, \allowbreak 0110, 1001, \dots, 0\bold{1}^x0, \allowbreak 1\bold{0}^x1\}$, where we denote a run of $r$ consecutive $0$'s (resp., $1$'s) as $\bold{0}^r$ (resp. $\bold{1}^r$) for brevity. Note that the set $\mathcal{S}_x$ is closed under taking pattern complements and has size $2x$. We define a binary symmetric $\mathcal{S}_x$-constrained code to be a code that forbids any pattern in the set $\mathcal{S}_x$ from appearing in any of its codewords. $\mathcal{S}_x$-constrained codes are used with bipolar non-return-to-zero (NRZ) signaling to control the separation between consecutive transitions. In NRZ signaling, a $0$ is represented by level $-A$ or the erasure level $E$ in Flash, while a $1$ is represented by level $+A$. Many applications, e.g., optical recording, require constrained codes to be balanced \cite{immink_opt}; a frequency domain approach is in  \cite{robert_spec1}.

In Flash systems employing block erasure, the error-prone patterns, i.e., the patterns that contribute the most to inter-cell interference (ICI), for NRZ signaling are somewhat different. It was demonstrated in \cite{qin_flash} that for Flash memories, the pattern $(q-1)0(q-1)$ should be eliminated, where $q$ is the number of levels in the cell and also the Galois field (GF) size\footnote{We directly map the GF elements to consecutive integers representing threshold voltage levels in order to follow the references.}. Balanced and constant-weight codes were designed to eliminate this pattern in \cite{qin_flash} and \cite{kayser_flash}, respectively. The authors of \cite{veeresh_mlc} showed that the set of patterns to eliminate in multi-level cell (MLC) Flash memories is $\{303, 313, 323\}$, which can be generalized to $\{(q-1)0(q-1), (q-1)1(q-1), \dots, (q-1)(q-2)(q-1)\}$ for $q$-level cell Flash memories as shown in \cite{chee_qlc}. The problem originates from the phenomenon that increasing the charge at the outer two cells causes the middle cell to have its charge also increased unintentionally because of the parasitic capacitances. Prior work only considered adjacent cells \cite{qin_flash, kayser_flash, veeresh_mlc, chee_qlc}. However, parasitic capacitances may also result in charge propagation between non-adjacent cells, which means patterns like $(q-1)\bold{0}^x(q-1)$, $x > 1$, can also be problematic (investigated here in the binary case).

We define the set $\mathcal{A}_x \triangleq \{101, 1001, \dots, 1\bold{0}^x1\}$ and note that $\mathcal{A}_x$ has size $x$. We define a binary asymmetric $\mathcal{A}_x$-constrained code to be a code that forbids any pattern in the set $\mathcal{A}_x$ from appearing in any of its codewords. The code is said to be asymmetric because $\mathcal{A}_x$ is not closed under taking pattern complements. Here, NRZ signaling is adopted. For magnetic recording channels having the extended partial-response 4 (EPR4) target, the authors of \cite{siegel_const} showed that asymmetric $\mathcal{A}_1$-constrained codes (forbidding $\{101\}$) achieve the same performance as symmetric $\mathcal{S}_1$-constrained codes (forbidding $\{010, 101\}$) with $20\%$ rate increase.

The idea of constructing constrained codes based on lexicographic indexing (also called enumerative coding) was first presented in \cite{tang_bahl} for RLL codes. The framework introduced in \cite{cover_lex} inspired more recent developments such as \cite{immink_lex} and \cite{braun_lex}. These developments have the drawback that the code length needs to be large for rates approaching capacity, resulting in limited rate-complexity trade-off advantages. The asymmetric constrained codes in \cite{chee_qlc}, designed for Flash systems, are enumerative constant-composition codes with high rates and average encoding-decoding complexity. However, these codes are limited in the sense that only the effect of the two adjacent cells is taken into account, i.e., for a single-level cell (SLC) Flash memory device (inaccurate nomenclature; it is a single-bit cell), they are only $\mathcal{A}_1$-constrained codes.

Recently, we introduced a new family of symmetric constrained codes, which we named lexicographically-ordered $\mathcal{S}_x$-constrained codes (LOCO codes) \cite{ahh_loco}. LOCO codes are capacity-achieving, and they offer up to $10\%$ rate gain, with low complexity encoding-decoding, compared with practical RLL codes designed for the same goal. A combination of LOCO codes and spatially-coupled graph-based codes \cite{ahh_tit2} resulted in significant density gains with limited rate reduction. Balancing LOCO codes was also proved to result in the minimum penalty in code rate. See \cite{ahh_loco} for details.

In this paper, we propose and analyze a new family of asymmetric constrained codes, which we name asymmetric lexicographically-ordered $\mathcal{A}_x$-constrained codes (A-LOCO codes), that improve performance by eliminating the error-prone patterns in Flash memories. Our A-LOCO codes can be constructed, encoded, and decoded for any set $\mathcal{A}_x$, making them capable of taking into account the effect of non-adjacent cells when needed. A-LOCO codes are capacity-achieving codes, and we establish a mapping-demapping formula between the lexicographic index and the codeword in order to enable simple, practical encoding and decoding algorithms. Compared with other practical asymmetric and symmetric constrained codes designed for the same purpose, A-LOCO codes offer higher rates at low complexity. In this paper, we only consider binary asymmetric constrained codes for SLC Flash memories. However, we expect to be able to develop non-binary asymmetric constrained codes for Flash memories with $q=2^y$, $y \geq 2$, levels. High rate non-binary asymmetric constrained codes will encourage the development of quad-level cell (QLC) Flash memories.

The rest of the paper is organized as follows. In Section~\ref{sec_card}, we enumerate the codewords in an A-LOCO code. In Section~\ref{sec_rule}, we establish the encoding-decoding rule of A-LOCO codes. In Section~\ref{sec_rate}, we discuss bridging, self-clocking, and rates of A-LOCO codes. In Section~\ref{sec_algo}, we introduce the algorithms, complexity analysis, and comparisons with other constrained codes. We conclude the paper in Section~\ref{sec_conc}.

%%%%%%%%%%%%%%%%%%%%%%%%%%%%%%%%%%%%%
\section{Cardinality of A-LOCO Codes}\label{sec_card}

In this section, we formally define A-LOCO codes, and then derive a recursive relation that gives the cardinality, i.e., the number of codewords, of these codes.

\begin{definition}\label{def_aloco}
An A-LOCO code $\mathcal{AC}_{m,x}$, with parameters $m \geq 1$ and $x \geq 1$, is defined via the following properties:
\begin{enumerate}
\item Codewords in $\mathcal{AC}_{m,x}$ are binary and of length $m$.
\item Codewords in $\mathcal{AC}_{m,x}$ are ordered lexicographically.
\item Any pattern in the asymmetric set $\mathcal{A}_x$ does not appear in any codeword $\bold{c}$ in $\mathcal{AC}_{m,x}$, where:
\begin{equation}\label{eqn_seta}
\mathcal{A}_x \triangleq \{101, 1001, \dots, 1\bold{0}^x1\}.
\end{equation}
\item The code $\mathcal{AC}_{m,x}$ contains all the codewords satisfying the previous three properties.
\end{enumerate}
\end{definition}

Lexicographic ordering of codewords means that they are ordered in an ascending manner following the rule $0 < 1$ for any bit, and the bit significance reduces from left to right.

Our main application in this work is Flash memories. In SLC devices, a $1$ results in a programmed cell, while a $0$ results in an unprogrammed cell, i.e., NRZ signaling. Thus, patterns of the form $1\bold{0}^x1$ are error-prone since they give rise to ICI on the inner cell(s) (the unprogrammed cell(s)).

Table~\ref{table_1} shows the codewords of the A-LOCO codes $\mathcal{AC}_{m,1}$, $m \in \{1, 2, \dots, 5\}$. The table demonstrates that an A-LOCO code with $m \notin \{1, 2\}$ is not closed under taking codeword complements. Moreover, the table also exhibits the increase in the A-LOCO code cardinality compared with the corresponding LOCO code. For example, the cardinality of the A-LOCO code with $m=5$ and $x=1$ is $21$, while it is only $16$ for the corresponding LOCO code \cite{ahh_loco}.

Next, we introduce a group structure for A-LOCO codes that helps us not only derive the cardinality recursively, but also devise the encoding-decoding rule of A-LOCO codes, which is based on lexicographic indexing.

For $m \geq 2$, the codewords in an A-LOCO code $\mathcal{AC}_{m,1}$ are classified into the following three groups:

\textbf{Group~1:} Codewords in this group start with $0$ from the left, i.e., at the left-most bit (LMB).

\textbf{Group~2:} Codewords in this group start with $11$ from the left, i.e., at the LMBs.

\textbf{Group~3:} Codewords in this group start with $1\bold{0}^{x+1}$ from the left, i.e., at the LMBs\footnote{In Group~3 and with $2 \leq m \leq x+1$, there exists only a single codeword, which has fewer bits than these LMBs, in the group. The following analysis also applies for such codewords.}.

This group structure is shown explicitly in Table~\ref{table_1} for $\mathcal{AC}_{5,1}$. Additionally, the horizontal lines in each column of codewords separate different groups. Note that bridging bits/symbols are required in order to guarantee that the forbidden patterns do not appear in streams of consecutive A-LOCO codewords. Bridging will be discussed later.

\begin{table*}
\caption{The codewords of five A-LOCO codes, $\mathcal{AC}_{m,1}$, $m \in \{1, 2, \dots, 5\}$ and $x=1$. The three different groups of codewords are shown for the code $\mathcal{AC}_{5,1}$.}
\vspace{-0.5em}
\centering
\scalebox{1.00}
{
\begin{tabular}{|c|c|c|c|c|c c|}
\hline
\multirow{2}{*}{Codeword index $g(\bold{c})$} & \multicolumn{6}{|c|}{\makecell{Codewords of the code $\mathcal{AC}_{m,1}$}} \\
\cline{2-7}
{} & \makecell{$m=1$} & \makecell{$m=2$} & \makecell{$m=3$} & \makecell{$m=4$} & \multicolumn{2}{|c|}{\makecell{$m=5$}} \\
\hline
$0$ & $0$ & $00$ & $000$ & $0000$ & $00000$ & \multirow{12}{*}{Group~1} \\
\cline{1-1}\cline{2-2}
$1$ & $1$ & $01$ & $001$ & $0001$ & $00001$ & \\
\cline{1-1}\cline{2-2}\cline{3-3}
$2$ &  & $10$ & $010$ & $0010$ & $00010$ & \\
\cline{1-1}\cline{3-3}
$3$ &  & $11$ & $011$ & $0011$ & $00011$ & \\
\cline{1-1}\cline{3-3}\cline{4-4}
$4$ &  &  & $100$ & $0100$ & $00100$ & \\
\cline{1-1}\cline{4-4}
$5$ &  &  & $110$ & $0110$ & $00110$ & \\
\cline{1-1}
$6$ &  &  & $111$ & $0111$ & $00111$ & \\
\cline{1-1}\cline{4-4}\cline{5-5}
$7$ &  &  &  & $1000$ & $01000$ & \\
\cline{1-1}
$8$ &  &  &  & $1001$ & $01001$ & \\
\cline{1-1}\cline{5-5}
$9$ &  &  &  & $1100$ & $01100$ & \\
\cline{1-1}
$10$ &  &  &  & $1110$ & $01110$ & \\
\cline{1-1}
$11$ &  &  &  & $1111$ & $01111$ & \\
\cline{1-1}\cline{5-5}\cline{6-7}
$12$ &  &  &  &  & $10000$ & \multirow{4}{*}{Group~3} \\
\cline{1-1}
$13$ &  &  &  &  & $10001$ & \\
\cline{1-1}
$14$ &  &  &  &  & $10010$ & \\
\cline{1-1}
$15$ &  &  &  &  & $10011$ & \\
\cline{1-1}\cline{6-7}
$16$ &  &  &  &  & $11000$ & \multirow{5}{*}{Group~2} \\
\cline{1-1}
$17$ &  &  &  &  & $11001$ & \\
\cline{1-1}
$18$ &  &  &  &  & $11100$ & \\
\cline{1-1}
$19$ &  &  &  &  & $11110$ & \\
\cline{1-1}
$20$ &  &  &  &  & $11111$ & \\
\hline
Code cardinality & $N(1, 1) \triangleq 2$  & $N(2, 1) = 4$ & $N(3, 1) = 7$ & $N(4, 1) = 12$ & \multicolumn{2}{|c|}{$N(5, 1) = 21$} \\
\hline
\end{tabular}}
\label{table_1}
\end{table*}

Theorem~\ref{thm_card} derives the cardinality of A-LOCO codes.

\begin{theorem}\label{thm_card}
Denote the cardinality of an A-LOCO code $\mathcal{AC}_{m,x}$ by $N(m, x)$ with:
\begin{equation}\label{eqn_Ndef}
N(m, x) \triangleq 1, \textup{ } m \leq 0, \textit{ and } N(1, x) \triangleq 2.
\end{equation}
The following recursive equation gives $N(m, x)$:
\begin{align}\label{eqn_Nrec}
N(m, x) &= 2N(m-1, x) - N(m-2, x) \nonumber \\ &+ N(m-x-2, x), \textup{ } m \geq 2.
\end{align}
\end{theorem}

\begin{IEEEproof}
We use the group structure illustrated above in order to prove Theorem~\ref{thm_card}.

\textbf{Group~1:} Each codeword from Group~1 in $\mathcal{AC}_{m,x}$ corresponds to a codeword in $\mathcal{AC}_{m-1,x}$ that shares the $m-1$ right-most bits (RMBs) with the codeword in $\mathcal{AC}_{m,x}$. This applies to all the codewords in $\mathcal{AC}_{m-1,x}$. Recall that patterns of the form $0\bold{1}^y0$, $1 \leq y \leq x$, are not forbidden in A-LOCO codes. Thus, the cardinality of Group~1 in $\mathcal{AC}_{m,x}$ is:
\begin{equation}\label{eqn_card1}
N_1(m, x) = N(m-1, x).
\end{equation}

\textbf{Group~2:} Each codeword from Group~2 in $\mathcal{AC}_{m,x}$ corresponds to a codeword in $\mathcal{AC}_{m-1,x}$ that starts with $1$ from the left and shares the $m-2$ RMBs with the codeword in $\mathcal{AC}_{m,x}$. This applies to all the codewords starting with $1$ from the left in $\mathcal{AC}_{m-1,x}$. Thus, the cardinality of Group~2 in $\mathcal{AC}_{m,x}$ is:
\begin{equation}\label{eqn_proof11}
N_2(m, x) = N(m-1, x)-N_1(m-1, x).
\end{equation}
Using (\ref{eqn_card1}) to compute $N_1(m-1, x)$ gives:
\begin{equation}\label{eqn_card2}
N_2(m, x) = N(m-1, x)-N(m-2, x).
\end{equation}

\textbf{Group~3:} Each codeword from Group~3 in $\mathcal{AC}_{m,x}$ corresponds to a codeword in $\mathcal{AC}_{m-x-1,x}$ that starts with $0$ from the left and shares the $m-x-2$ RMBs with the codeword in $\mathcal{AC}_{m,x}$. This applies to all the codewords starting with $0$ from the left in $\mathcal{AC}_{m-x-1,x}$. Thus, the cardinality of Group~3 in $\mathcal{AC}_{m,x}$ is:
\begin{equation}\label{eqn_proof12}
N_3(m, x) = N_1(m-x-1, x).
\end{equation}
Using (\ref{eqn_card1}) to compute $N_1(m-x-1, x)$ gives:
\begin{equation}\label{eqn_card3}
N_3(m, x) = N(m-x-2, x).
\end{equation}

Adding (\ref{eqn_card1}), (\ref{eqn_card2}), and (\ref{eqn_card3}) gives:
\begin{align}
&N(m, x) = \sum_{\ell = 1}^{3} N_\ell(m, x) \nonumber \\ &= 2N(m-1, x) - N(m-2, x) + N(m-x-2, x), \nonumber
\end{align}
which completes the proof.
\end{IEEEproof}

\begin{example}\label{ex_1}
The cardinalities of $\mathcal{AC}_{m,1}$, $m \in \{2, 3, 4, 5\}$, are computed using Theorem~\ref{thm_card} as follows:
\begin{align}
&N(-1, 1) \triangleq 1, \textup{ }  N(0, 1) \triangleq 1, \textup{ } N(1, 1) \triangleq 2, \nonumber \\
&N(2, 1) = 2N(1, 1) - N(0, 1) + N(-1, 1) = 4, \nonumber \\
&N(3, 1) = 2N(2, 1) - N(1, 1) + N(0, 1) = 7, \allowdisplaybreaks \nonumber \\
&N(4, 1) = 2N(3, 1) - N(2, 1) + N(1, 1) = 12, \nonumber \\
&N(5, 1) = 2N(4, 1) - N(3, 1) + N(2, 1) = 21, \nonumber
\end{align}
which are also given in the last row of Table~\ref{table_1}. The cardinalities of the three groups can be used to compute $N(5, 1)$, for example, as follows:
\begin{align}
N_1(5, 1) &= N(4, 1) = 12, \nonumber \\
N_2(5, 1) &= N(4, 1) - N(3, 1) = 5, \nonumber \\
N_3(5, 1) &= N(2, 1) = 4, \nonumber \\
N(5, 1) &= N_1(5, 1) + N_2(5, 1) + N_3(5, 1) = 21. \nonumber
\end{align}
These numbers are also consistent with the groups shown in the last column of Table~\ref{table_1}.
\end{example}

Theorem~\ref{thm_card} is important because, for a given length $m$, the number of codewords determines the rate of the code. This is true for all coding techniques based on lexicographic ordering (true for enumerative coding techniques in general). Theorem~\ref{thm_card} is also essential for devising the encoding-decoding rule of A-LOCO codes as we shall see next section, and consequently, the encoding-decoding algorithms.

%%%%%%%%%%%%%%%%%%%%%%%%%%%%%%%%%%%%%
\vspace{+0.2em}
\section{Encoding-Decoding Rule of A-LOCO Codes}\label{sec_rule}

In this section, we devise the encoding-decoding rule of A-LOCO codes, which is based on lexicographic indexing. This rule is what enables simple, low complexity encoding and decoding algorithms for A-LOCO codes.

We first introduce some notation. Define an A-LOCO codeword of length $m$ as $\bold{c} \triangleq [c_{m-1} \textup{ } c_{m-2} \textup{ } \dots \textup{ } c_0 ] \in \mathcal{AC}_{m,x}$. The index of an A-LOCO codeword $\bold{c}$ in $\mathcal{AC}_{m,x}$ is denoted by $g(m, x, \bold{c})$, which is sometimes abbreviated to $g(\bold{c})$, as in Table~\ref{table_1}, for simplicity. We also define an integer variable $a_i$ for each binary $c_i$ as follows:
\vspace{-0.1em}\begin{align}\label{eqn_ai}
a_i \triangleq \left\{\begin{matrix}1, \textup{ } &c_i = 1,
\\ 0, \textup{ } &c_i = 0,
\end{matrix}\right.
\end{align}
with $a_m \triangleq 0$. The same notation applies for an A-LOCO codeword of length $m+1$, $\bold{c}'$ in $\mathcal{AC}_{m+1,x}$, and an A-LOCO codeword of length $m-x$, $\bold{c}''$ in $\mathcal{AC}_{m-x,x}$.

Theorem~\ref{thm_rule} gives the encoding-decoding rule of A-LOCO codes. The indexing is trivial for the case of $m=1$

\vspace{+0.3em}\begin{theorem}\label{thm_rule}
The lexicographic index $g(m, x, \bold{c})$ of an A-LOCO codeword $\bold{c}$ in $\mathcal{AC}_{m,x}$, $m \geq 2$ and $x \geq 1$, is computed from the codeword itself according to the following rule:
\begin{equation}\label{eqn_rule}
g(\bold{c}) = \sum_{i=0}^{m-1} a_i N(i-a_{i+1}x, x).
\end{equation}
\end{theorem}

\begin{IEEEproof}
We prove Theorem~\ref{thm_rule} by induction.

\textbf{Base:} Our base case is the case of $m=2$. For $\mathcal{AC}_{2,x}$, we always have four codewords, say $\bold{c}_0$, $\bold{c}_1$, $\bold{c}_2$, and $\bold{c}_3$ in order, for any value of $x$. These four codewords are listed in Table~\ref{table_1}. We want to prove that $g(\bold{c}_j) = j$, for all $j \in \{0, 1, 2, 3\}$, using (\ref{eqn_rule}). The bits of a codeword $\bold{c}_j$ are $c_{j, i}$, $i \in \{0, 1\}$, and $a_{j, i}$ is defined for each $c_{j, i}$ as in (\ref{eqn_ai}).
\begin{align}\label{eqn_base}
g(\bold{c}_0) &= \sum_{i=0}^{1} 0 \cdot N(i-a_{i+1}x, x) = 0, \nonumber \\
g(\bold{c}_1) &= \sum_{i=0}^{1} a_i N(i-a_{i+1}x, x) = N(0-0, x) = 1, \nonumber \\
g(\bold{c}_2) &= \sum_{i=0}^{1} a_i N(i-a_{i+1}x, x) = N(1-0, x) = 2, \nonumber \\
g(\bold{c}_3) &= \sum_{i=0}^{1} a_i N(i-a_{i+1}x, x) \nonumber \\ &= N(1-0, x) + N(0-1, x) = 2 + 1 = 3.
\end{align}
Recall that $N(1, x) \triangleq 2$, for all $x \in \{1, 2, \dots\}$.

\textbf{Assumption:} We assume that the following is correct:
\begin{equation}\label{eqn_assume}
g(\overline{m}, x, \overline{\bold{c}}) = \sum_{i=0}^{\overline{m}-1} \overline{a}_i N(i-\overline{a}_{i+1}x, x),
\end{equation}
where $\overline{\bold{c}} \in \mathcal{AC}_{\overline{m},x}$ and $\overline{m} \in \{2, 3, \dots, m\}$ with the same notation defined before Theorem~\ref{thm_rule} applied to $\overline{\bold{c}}$. The assumption here basically means (\ref{eqn_rule}) is correct for all A-LOCO codes $\mathcal{AC}_{\overline{m},x}$, $\overline{m} \in \{1, 2, \dots, m\}$.

\textbf{To be proved:} We prove that:
\begin{equation}\label{eqn_tbpr}
g(m+1, x, \bold{c}') = \sum_{i=0}^{m} a'_i N(i-a'_{i+1}x, x),
\end{equation}
which means we prove that given the base and the assumption, (\ref{eqn_rule}) is also correct for the A-LOCO code $\mathcal{AC}_{m+1,x}$.

One more time, we use the group structure introduced in Section~\ref{sec_card} to prove (\ref{eqn_tbpr}). Note that the group structure can be defined for $\mathcal{AC}_{m+1,x}$ as defined for $\mathcal{AC}_{m,x}$. We use the same codeword correspondence in the proof of Theorem~\ref{thm_card} for the three groups (with $m+1$ replacing $m$).

\textbf{Group~1:} The codewords in Group~1 in $\mathcal{AC}_{m+1,x}$ start at index $0$, and the corresponding codewords in $\mathcal{AC}_{m,x}$ also start at index $0$. Thus, for this group, the shift in codeword indices between a codeword $\bold{c}'$ in $\mathcal{AC}_{m+1,x}$ and the corresponding codeword $\bold{c}$ in $\mathcal{AC}_{m,x}$ is:
\begin{equation}\label{eqn_shift1}
g(m+1, x, \bold{c}') - g(m, x, \bold{c}) = 0.
\end{equation}
Consequently, and using (\ref{eqn_assume}):
\begin{equation}\label{eqn_proof21}
g(m+1, x, \bold{c}') = g(m, x, \bold{c}) = \sum_{i=0}^{m-1} a_i N(i-a_{i+1}x, x).
\end{equation}
Since $\bold{c}'$ starts with $0$ from the left, $a'_m = 0$. Additionally, $\bold{c}'$ and $\bold{c}$ share the $m$ RMBs. Thus, (\ref{eqn_proof21}) can be written as:
\begin{equation}\label{eqn_rule1}
g(m+1, x, \bold{c}') = \sum_{i=0}^{m} a'_i N(i-a'_{i+1}x, x).
\end{equation}

\textbf{Group~2:} The codewords in Group~2 in $\mathcal{AC}_{m+1,x}$ start right after Groups~1 and 3, and the corresponding codewords in $\mathcal{AC}_{m,x}$ start right after all the codewords that start with $0$ from the left. Thus, for this group, the shift in codeword indices between a codeword $\bold{c}'$ in $\mathcal{AC}_{m+1,x}$ and the corresponding codeword $\bold{c}$ in $\mathcal{AC}_{m,x}$ is:
\begin{align}\label{eqn_shift2}
&g(m+1, x, \bold{c}') - g(m, x, \bold{c}) \nonumber \\ &= N_1(m+1, x) + N_3(m+1, x) - N_1(m, x) \nonumber \\ &= N(m, x) + N(m-x-1, x) - N(m-1, x),
\end{align}
where the second equality follows from using (\ref{eqn_card1}) and (\ref{eqn_card3}). Consequently, and using (\ref{eqn_assume}):
\begin{align}\label{eqn_proof22}
&g(m+1, x, \bold{c}') \nonumber \\ &= N(m, x) + N(m-x-1, x) - N(m-1, x) \nonumber \\ &+ \sum_{i=0}^{m-1} a_i N(i-a_{i+1}x, x).
\end{align}
Observe that $c_{m-1} = 1$; thus, $a_{m-1} = 1$, while $a_m \triangleq 0$. Using these observations in (\ref{eqn_proof22}) results in:
\begin{align}\label{eqn_proof23}
&g(m+1, x, \bold{c}') \nonumber \\ &= N(m, x) + N(m-x-1, x) - N(m-1, x) \nonumber \\ &+ N(m-1, x) + \sum_{i=0}^{m-2} a_i N(i-a_{i+1}x, x).
\end{align}
Since $\bold{c}'$ starts with $11$ from the left, $a'_m = a'_{m-1} = 1$, same as $a_{m-1}$, while $a'_{m+1} \triangleq 0$. Consequently,
\begin{align}\label{eqn_proof24}
&N(m, x) + N(m-x-1, x) \nonumber \\ &= a'_m N(m - a'_{m+1}x, x) + a'_{m-1} N(m -1 - a'_m x, x) \nonumber \\ &= \sum_{i=m-1}^{m} a'_i N(i-a'_{i+1}x, x).
\end{align}
Additionally, $\bold{c}'$ and $\bold{c}$ share the $m-1$ RMBs. Thus, aided by (\ref{eqn_proof24}), (\ref{eqn_proof23}) can be written as:
\vspace{-0.1em}\begin{align}\label{eqn_rule2}
&g(m+1, x, \bold{c}') \nonumber \\ &= \sum_{i=m-1}^{m} a'_i N(i-a'_{i+1}x, x) + \sum_{i=0}^{m-2} a'_i N(i-a'_{i+1}x, x) \nonumber \\ &= \hspace{+0.6em} \sum_{i=0}^{m} a'_i N(i-a'_{i+1}x, x).
\end{align}

\textbf{Group~3:} The codewords in Group~3 in $\mathcal{AC}_{m+1,x}$ start right after Group~1, and the corresponding codewords in $\mathcal{AC}_{m-x,x}$ start at index $0$. Thus, for this group, the shift in codeword indices between a codeword $\bold{c}'$ in $\mathcal{AC}_{m+1,x}$ and the corresponding codeword $\bold{c}''$ in $\mathcal{AC}_{m-x,x}$ is:
\begin{align}\label{eqn_shift3}
&g(m+1, x, \bold{c}') - g(m-x, x, \bold{c}'') \nonumber \\ &= N_1(m+1, x) = N(m, x).
\end{align}
Consequently, and using (\ref{eqn_assume}):
\begin{align}\label{eqn_proof25}
&g(m+1, x, \bold{c}') = N(m, x) + \sum_{i=0}^{m-x-1} a''_i N(i-a''_{i+1}x, x) \nonumber \\ &=N(m, x) + \sum_{i=0}^{m-x-2} a''_i N(i-a''_{i+1}x, x),
\end{align}
where the second equality follows from that $\bold{c}''$ starts with $0$ from the left; thus, $a''_{m-x-1} = 0$. Since $\bold{c}'$ starts with $1\bold{0}^{x+1}$ from the left, $a'_m = 1$ and $a'_{m-1}= a'_{m-2} = \dots = a'_{m-x-1} = 0$, while $a'_{m+1} \triangleq 0$. Additionally, $\bold{c}'$ and $\bold{c}''$ share the $m-x-1$ RMBs. Thus, (\ref{eqn_proof25}) can be written as:
\begin{equation}\label{eqn_rule3}
g(m+1, x, \bold{c}') = \sum_{i=0}^{m} a'_i N(i-a'_{i+1}x, x).
\end{equation}

From (\ref{eqn_rule1}), (\ref{eqn_rule2}), and (\ref{eqn_rule3}), (\ref{eqn_tbpr}) is proved, which completes the proof by induction for any A-LOCO code $\mathcal{AC}_{m,x}$, $m \geq 2$ and $x \geq 1$.
\end{IEEEproof}

\begin{example}\label{ex_2}
We illustrate Theorem~\ref{thm_rule} by applying (\ref{eqn_rule}) on two different codewords in $\mathcal{AC}_{5,1}$, given in Table~\ref{table_1}, to compute their indices. The first codeword is $01111$; thus, $a_4 = 0$ and $a_3 = a_2 = a_1 = a_0 = 1$. Consequently,
\begin{align}
&g(\bold{c}) = \sum_{i=0}^{4} a_i N(i-a_{i+1}, 1) \nonumber \\ &= N(3-0, 1) + N(2-1, 1) + N(1-1, 1) + N(0-1, 1) \nonumber \\ &= N(3, 1) + N(1, 1) + N(0, 1) + N(-1, 1) \nonumber \\ &= 7+2+1+1 = 11, \nonumber
\end{align}
which is indeed the index of this codeword in Table~\ref{table_1}. The second codeword is $11001$; thus, $a_4 = a_3 = a_0 = 1$ and $a_2 = a_1 = 0$, while $a_5 \triangleq 0$. Consequently,
\begin{align}
&g(\bold{c}) = \sum_{i=0}^{4} a_i N(i-a_{i+1}, 1) \nonumber \\ &= N(4-0, 1) + N(3-1, 1) + N(0-0, 1) \nonumber \\ &= N(4, 1) + N(2, 1) + N(0, 1) \nonumber \\ &= 12+4+1 = 17, \nonumber
\end{align}
which is indeed the index of this codeword in Table~\ref{table_1}.
\end{example}

Theorem~\ref{thm_rule} gives the encoding-decoding rule of A-LOCO codes. In particular, this theorem provides a simple mapping-demapping (both are one-to-one) from the index to the codeword and vice-versa. This simple mapping-demapping represented by (\ref{eqn_rule}) is what enables low-complexity encoding and decoding algorithms for A-LOCO codes as we shall see later. The main advantage offered by A-LOCO codes is that they are capacity-achieving asymmetric constrained codes with simple encoding-decoding.

%%%%%%%%%%%%%%%%%%%%%%%%%%%%%%%%%%%%%
\section{Bridging, Clocking, and Achievable Rates}\label{sec_rate}

In this section, we discuss the bridging patterns and the self-clocking of A-LOCO codes. Then, we introduce the rates of A-LOCO codes in the finite-length regime and show that they are capacity-achieving codes.

In fixed-length constrained codes \cite{immink_lex}, given any two consecutive codewords, bridging patterns are needed to prevent forbidden patterns from appearing on the transition from the first codeword to the following codeword \cite{ahh_loco}. For example, consider the A-LOCO code $\mathcal{AC}_{5,1}$ given in Table~\ref{table_1}, if the codewords having indices $13$ and $8$ are to be written consecutively without bridging, we get the following substream of bits $1000\textit{101}001$ in which, the forbidden pattern $101$ does appear (the pattern is shown in italic).

For symmetric LOCO codes, it was shown in \cite{ahh_loco} that upon deciding the bridging method, there is a compromise between the maximum protection of the bits at the codeword transitions and the minimum number of bits/symbols to be used for bridging. The optimal bridging method for a symmetric LOCO code with parameter $x$ in terms of bits protection was shown to require $2x$ bridging bits, which results in significant rate loss \cite{ahh_loco}. Thus, we adopted a suboptimal bridging method for symmetric LOCO codes that is bridging by $x$ no writing, or no transmission, symbols.

For an A-LOCO code with parameter $x$, only patterns~of the form $\{101, 1001, \dots, 1\bold{0}^x1\}$ are forbidden. Thus, bridging with $x$ $0$'s, i.e., with the pattern $\bold{0}^x$, ensures that forbidden patterns do not appear on the codeword transitions except for the case when the RMB of a codeword and the LMB of the next codeword to be written are both $1$'s. In this case, bridging with $x$ $1$'s, i.e., with the pattern $\bold{1}^x$, is used instead. In summary, our bridging method for A-LOCO codes is:
\begin{enumerate}
\item If the RMB of a codeword and the LMB of the next codeword to be written are both $1$'s, bridge with $\bold{1}^x$.
\item Otherwise, bridge with $\bold{0}^x$.
\end{enumerate}

It is important here to highlight that the proposed bridging methods is one of the advantages of A-LOCO codes over other constrained codes. In particular, the bridging method is not only optimal in terms of bits protection, but also requires the minimum number of added bits for bridging, which is only $x$ bits for an A-LOCO code $\mathcal{AC}_{m,x}$. Observe that this bridging method is also easy to implement.

Next, we discuss the self-clocking of A-LOCO codes. This feature is required in order to have clock recovery and system calibration \cite{siegel_mr, ahh_loco}. A constrained code is said to be self-clocked if any stream of codewords to be written contains a sufficient number of appropriately-separated transitions after bridging and signaling are applied. Since NRZ signaling is adopted for A-LOCO codes, these transitions are the $0-1$ and $1-0$ transitions in A-LOCO codewords. To achieve self-clocking, we just need to remove the two codewords $\bold{0}^m$ and $\bold{1}^m$ from an A-LOCO code $\mathcal{AC}_{m,x}$ as this guarantees at least one transition in each A-LOCO codeword.

\begin{definition}\label{def_caloco}
A self-clocked A-LOCO code (CA-LOCO code) $\mathcal{AC}^{\textup{c}}_{m,x}$, $m \geq 2$, is the A-LOCO code $\mathcal{AC}_{m,x}$ after removing the all $0$'s and the all $1$'s codewords. Mathematically,
\begin{equation}\label{eqn_caloco}
\mathcal{AC}^{\textup{c}}_{m,x} \triangleq \mathcal{AC}_{m,x} \setminus \{\bold{0}^m, \bold{1}^m\}.
\end{equation}
Thus, the cardinality of $\mathcal{AC}^{\textup{c}}_{m,x}$ is:
\begin{equation}\label{eqn_caloco}
N^{\textup{c}}(m, x) = N(m, x)-2.
\end{equation}
\end{definition}

We define $k^{\textup{c}}_{\textup{eff}}$ as the maximum number of consecutive cells between two consecutive transitions (all programmed or all unprogrammed) after a stream of CA-LOCO codewords separated by bridging patterns is written; one bit per cell. Thus, $k^{\textup{c}}_{\textup{eff}}$ is the length of the longest run of consecutive $1$'s or $0$'s in a stream of CA-LOCO codewords separated by bridging patterns. The scenarios under which $k^{\textup{c}}_{\textup{eff}}$ is achieved are:
\begin{align}\label{eqn_scen}
&1\bold{0}^{m-1} - \bold{0}^x - \bold{0}^{m-1}1 \textup{ } \textup{ and} \nonumber \\
&0\bold{1}^{m-1} - \bold{1}^x - \bold{1}^{m-1}0. \nonumber
\end{align}
Consequently, $k^{\textup{c}}_{\textup{eff}}$ is given by:
\begin{equation}\label{eqn_keff}
k^{\textup{c}}_{\textup{eff}} = 2(m-1)+x,
\end{equation}
which is the same equation satisfied by LOCO codes \cite{ahh_loco}.

Given the cardinality of a CA-LOCO code $\mathcal{AC}^{\textup{c}}_{m,x}$, the size of the messages $\mathcal{AC}^{\textup{c}}_{m,x}$ encodes is:
\begin{equation}\label{eqn_sc}
s^{\textup{c}} = \left \lfloor \log_2 N^{\textup{c}}(m, x) \right \rfloor = \left \lfloor \log_2 (N(m, x)-2) \right \rfloor.
\end{equation}
Consequently, the rate of a CA-LOCO code $\mathcal{AC}^{\textup{c}}_{m,x}$, where $x$ bits are used for bridging as illustrated above, is given by:
\begin{equation}\label{eqn_rate}
R^{\textup{c}}_{\textup{A-LOCO}} = \frac{s^{\textup{c}}}{m+x} = \frac{\left \lfloor \log_2 (N(m, x)-2) \right \rfloor}{m+x}.
\end{equation}

Observe the following:
\begin{enumerate}
\item A CA-LOCO code $\mathcal{AC}^{\textup{c}}_{m,x}$ contains all the codewords satisfying the $\mathcal{A}_x$ constraint except the two codewords in $\{\bold{0}^m, \bold{1}^m\}$. This follows from Definitions~\ref{def_aloco} and \ref{def_caloco}.

\item The number of bits added for bridging is $x$, which does not grow with the code length $m$. Thus, as $m \rightarrow \infty$, the $x$ in the denominator of (\ref{eqn_rate}) can be ignored.
\end{enumerate}
From the above two observations, we conclude that CA-LOCO codes are \textbf{capacity-achieving constrained codes}. Shortly, we will show that rates approaching the capacity can be achieved with low complexity encoding-decoding.

\begin{table}
\caption{The codewords of the CA-LOCO code $\mathcal{AC}^{\textup{c}}_{5,1}$ and the corresponding messages.}
\vspace{-0.5em}
\centering
\scalebox{1.00}
{
\begin{tabular}{|c|c|c|}
\hline
\makecell{Message $\bold{b}$} & \makecell{Index $g(\bold{c})$} & \makecell{Codeword $\bold{c}$} \\
\hline
$0000$ & $1$ & $00001$ \\
\hline
$0001$ & $2$ & $00010$ \\
\hline
$0010$ & $3$ & $00011$ \\
\hline
$0011$ & $4$ & $00100$ \\
\hline
$0100$ & $5$ & $00110$ \\
\hline
$0101$ & $6$ & $00111$ \\
\hline
$0110$ & $7$ & $01000$ \\
\hline
$0111$ & $8$ & $01001$ \\
\hline
$1000$ & $9$ & $01100$ \\
\hline
$1001$ & $10$ & $01110$ \\
\hline
$1010$ & $11$ & $01111$ \\
\hline
$1011$ & $12$ & $10000$ \\
\hline
$1100$ & $13$ & $10001$ \\
\hline
$1101$ & $14$ & $10010$ \\
\hline
$1110$ & $15$ & $10011$ \\
\hline
$1111$ & $16$ & $11000$ \\
\hline
\end{tabular}}
\label{table_2}
\end{table}

\begin{example}\label{ex_3}
Consider again the A-LOCO code $\mathcal{AC}_{5,1}$ in Table~\ref{table_1}. The CA-LOCO code $\mathcal{AC}^{\textup{c}}_{5,1}$ is obtained by removing the codewords $\bold{0}^m$ (with index $0$) and $\bold{1}^m$ (with index $20$) from $\mathcal{AC}_{5,1}$. For this CA-LOCO code, we have:
\begin{equation}\label{eqn_ex31}
k^{\textup{c}}_{\textup{eff}} = 2(5-1)+1 = 9. \nonumber
\end{equation}
The size of the messages $\mathcal{AC}^{\textup{c}}_{5,1}$ encodes is:
\begin{equation}\label{eqn_ex32}
s^{\textup{c}} = \left \lfloor \log_2 (N(5, 1)-2) \right \rfloor = \left \lfloor \log_2 19 \right \rfloor = 4, \nonumber
\end{equation}
where $N(5, 1) = 21$ from Example~\ref{ex_1} and Table~\ref{table_1}. All the $16$ codewords of $\mathcal{AC}^{\textup{c}}_{5,1}$ that have corresponding messages are shown in Table~\ref{table_2}. From (\ref{eqn_rate}), the rate is:
\begin{equation}\label{eqn_ex33}
R^{\textup{c}}_{\textup{A-LOCO}} = \frac{4}{5+1} = 0.6667. \nonumber
\end{equation}
Note that this is a relatively low rate because of the small value of the code length $m$.
\end{example}

Table~\ref{table_3} lists the rates of multiple CA-LOCO codes $\mathcal{AC}^{\textup{c}}_{m,x}$ for different values of $m$ and $x \in \{1, 2\}$. For the case of $x=1$, at length $m=44$ (resp., $76$), the rate is $0.8000$ (resp., $0.8052$). The capacity of $\mathcal{A}_1$-constrained codes is $0.8114$ \cite{siegel_const, chee_qlc}. Thus, at length $76$ (resp., $113$) bits, the CA-LOCO code is within just $0.8\%$ (resp., $0.6\%$) from the capacity. In fact, for the larger length of $357$ bits, the CA-LOCO code is within only $0.2\%$ from the capacity. For the case of $x=2$, at length $m=28$ (resp., $64$), the rate is $0.6667$ (resp., $0.6818$). The capacity of $\mathcal{A}_2$-constrained codes is $0.6942$ from the finite-state transition diagram (FSTD). Thus, at length $64$ (resp., $123$) bits, the CA-LOCO code is within just $1.8\%$ (resp., $0.9\%$) from the capacity. In fact, for the larger length of $244$ bits, the CA-LOCO code is within only $0.5\%$ from the capacity.

\begin{remark}\label{rmk_abal}
A-LOCO codes do not satisfy the complement rule of symmetric LOCO codes in \cite[Lemma 3]{ahh_loco}. Thus, balancing A-LOCO codes incurs a higher rate penalty. To reduce this penalty, almost-balanced A-LOCO codes, where no strict guarantee exists on the maximum magnitude of the running disparity, can be designed in a way similar to balanced LOCO codes. The balancing requirement is more important in optical recording systems than it is in Flash systems.
\end{remark}

\begin{table}
\caption{Rates of CA-LOCO codes $\mathcal{AC}^{\textup{c}}_{m,x}$ for different values of $m$ and $x \in \{1, 2\}$.}
\vspace{-0.5em}
\centering
\scalebox{1.00}
{
\begin{tabular}{|c|c|c|}
\hline
\makecell{Code parameters} & \makecell{Rate} & \makecell{Adder size} \\
\hline
$m=17$ and $x=1$ & $0.7778$ & $14$ bits \\
\hline
$m=44$ and $x=1$ & $0.8000$ & $36$ bits \\
\hline
$m=76$ and $x=1$ & $0.8052$ & $62$ bits \\
\hline
$m=113$ and $x=1$ & $0.8070$ & $92$ bits \\
\hline
$m=357$ and $x=1$ & $0.8101$ &  \\
\hline
$m=18$ and $x=2$ & $0.6500$ & $13$ bits \\
\hline
$m=28$ and $x=2$ & $0.6667$ & $20$ bits \\
\hline
$m=64$ and $x=2$ & $0.6818$ & $45$ bits \\
\hline
$m=123$ and $x=2$ & $0.6880$ & $86$ bits \\
\hline
$m=244$ and $x=2$ & $0.6911$ &  \\
\hline
\end{tabular}}
\label{table_3}
\end{table}

%%%%%%%%%%%%%%%%%%%%%%%%%%%%%%%%%%%%%
\section{Algorithms, Complexity, and Comparisons}\label{sec_algo}

In this section, we introduce practical encoding and decoding algorithms of A-LOCO codes. We then discuss the complexity of these algorithms, and make comparisons with other asymmetric and symmetric constrained codes that mitigate ICI in Flash systems.

In coding techniques based on indexing points (here representing codewords), devising simple algorithms to perform the mapping-demapping between the index and the associated point is critical to avoid look-up tables; and thus, to make the technique practical for large sizes. For example, motivated by this observation, the authors of \cite{laroia_const} developed simple algorithms to index the points of multi-dimensional constellations. The simple, practical algorithms we introduce in this section are also motivated by the same observation.

Algorithm~\ref{alg_enc} is the encoding algorithm. While generating a specific codeword $\bold{c}$ in the algorithm, we define the RMB of the previous codeword as $\zeta_0$.

\begin{algorithm}
\caption{Encoding CA-LOCO Codes}
\begin{algorithmic}[1]
\State \textbf{Input:} Incoming stream of binary messages.
\State Decide the value of $x$ based on system requirements.
\State Use (\ref{eqn_Ndef}) and (\ref{eqn_Nrec}) to compute $N(i, x)$, $i \in \{2, 3, \dots\}$.
\State Specify $m$, the smallest $i$ in Step~3 to achieve the desired rate. Then, $s^{\textup{c}} = \left \lfloor \log_2 \left( N(m, x) - 2 \right )  \right \rfloor$.
\State \textbf{for} each incoming message $\bold{b}$ of length $s^{\textup{c}}$ \textbf{do}
\State \hspace{2ex} Compute $g(\bold{c})=\textup{decimal}(\bold{b})+1$. \textit{(binary sequence to decimal integer)}
\State \hspace{2ex} Initialize $\textup{residual}$ with $g(\bold{c})$ and $c_m$ with $0$.
\State \hspace{2ex} \textbf{for} $i \in \{m-1, m-2, \dots, 0\}$ \textbf{do} \textit{(in order)}
\State \hspace{4ex} \textbf{if} $c_{i+1} = 0$ \textbf{then}
\State \hspace{6ex} Set $\textup{subt\_index} = i$.
\State \hspace{4ex} \textbf{else}
\State \hspace{6ex} Set $\textup{subt\_index} = i - x$.
\State \hspace{4ex} \textbf{end if}
\State \hspace{4ex} \textbf{if} $\textup{residual} < N(\textup{subt\_index}, x)$ \textbf{then}
\State \hspace{6ex} Encode $c_i = 0$.
\State \hspace{4ex} \textbf{else}
\State \hspace{6ex} Encode $c_i = 1$.
\State \hspace{6ex} $\textup{residual} \leftarrow \textup{residual} - N(\textup{subt\_index}, x)$.
\State \hspace{4ex} \textbf{end if}
\State \hspace{4ex} \textbf{if} $i = m-1$ \textbf{then}
\State \hspace{6ex} \textbf{if} $\zeta_0 = 1$ and $c_{m-1} = 1$ \textbf{then}
\State \hspace{8ex} Bridge with $x$ $1$'s, i.e., $\bold{1}^x$, before $c_{m-1}$.
\State \hspace{6ex} \textbf{else}
\State \hspace{8ex} Bridge with $x$ $0$'s, i.e., $\bold{0}^x$, before $c_{m-1}$.
\State \hspace{6ex} \textbf{end if}
\State \hspace{4ex} \textbf{end if}
\State \hspace{2ex} \textbf{end for}
\State \textbf{end for}
\State \textbf{Output:} Outgoing stream of binary CA-LOCO codewords. \textit{(to be written on the SLC Flash device)}
\end{algorithmic}
\label{alg_enc}
\end{algorithm}

\begin{example}\label{ex_4}
We apply Algorithm~\ref{alg_enc} to encode the message $1010$ using the CA-LOCO code $\mathcal{AC}^{\textup{c}}_{5,1}$ ($m=5$ and $x=1$). Recall that $N(1, 1) \triangleq 2$, $N(2, 1)=4$, $N(3, 1)=7$, $N(4, 1)=12$, and $N(5, 1)=21$ (see Example~\ref{ex_1}). From Step~6, $g(\bold{c}) = \textup{decimal}(1010) + 1 = 11$, which is the initial value of $\textup{residual}$. The bits of the codeword $\bold{c}$ are generated as follows:
\begin{enumerate}
\item For $i=4$, and since $c_5$ is set to $0$, $\textup{subt\_index} = i = 4$ from Step~10. Now, $\textup{residual} = 11 < N(4, 1) = 12$. Thus, $c_4$ is encoded to $0$ from Step~15. Then, the \textbf{if} condition in Step~20 is satisfied, and because $c_4 \neq 1$, we bridge with $\bold{0}^x$ before $c_4$ assuming that this is not the first codeword.

\item For $i=3$, and since $c_4 = 0$, $\textup{subt\_index} = i = 3$ from Step~10. Now, $\textup{residual} = 11 > N(3, 1) = 7$. Thus, $c_3$ is encoded to $1$ from Step~17, and $\textup{residual}$ becomes $11-7=4$ from Step~18.

\item For $i=2$, and since $c_3 = 1$, $\textup{subt\_index} = i-x = 1$ from Step~12. Now, $\textup{residual} = 4 > N(1, 1) \triangleq 2$. Thus, $c_2$ is encoded to $1$ from Step~17, and $\textup{residual}$ becomes $4-2=2$ from Step~18.

\item For $i=1$, and since $c_2 = 1$, $\textup{subt\_index} = i-x = 0$ from Step~12. Now, $\textup{residual} = 2 > N(0, 1) \triangleq 1$. Thus, $c_1$ is encoded to $1$ from Step~17, and $\textup{residual}$ becomes $2-1=1$ from Step~18.

\item For $i=0$, and since $c_1 = 1$, $\textup{subt\_index} = i-x = -1$ from Step~12. Now, $\textup{residual} = 1 = N(-1, 1) \triangleq 1$. Thus, $c_0$ is encoded to $1$ from Step~17, and $\textup{residual}$ becomes $1-1=0$ from Step~18.
\end{enumerate}
As a result of this procedure, the message $1010$ is encoded using the CA-LOCO code $\mathcal{AC}^{\textup{c}}_{5,1}$ to the codeword $01111$, which is consistent with Table~\ref{table_2}.
\end{example}

Observe that Algorithm~\ref{alg_enc} has less steps and less computations compared with \cite[Algorithm~1]{ahh_loco} for symmetric LOCO codes. The reason is that all the steps required to avoid the patterns in $\{010, 0110, \dots, 0\bold{1}^x0\}$ are not needed here since these patterns are not forbidden for A-LOCO codes. Thus, the encoding complexity of an A-LOCO code is less than that of the LOCO code with the same $m$ and $x$.

Algorithm~\ref{alg_dec} is the decoding algorithm.

\begin{algorithm}
\caption{Decoding CA-LOCO Codes}
\begin{algorithmic}[1]
\State \textbf{Inputs:} Incoming stream of binary CA-LOCO codewords, in addition to $m$, $x$, and $s^{\textup{c}}$.
\State Use (\ref{eqn_Ndef}) and (\ref{eqn_Nrec}) to compute $N(i, x)$, $i \in \{2, 3, \dots, m\}$.
\State \textbf{for} each incoming codeword $\bold{c}$ of length $m$ \textbf{do}
\State \hspace{2ex} Initialize $g(\bold{c})$ with $0$ and $c_m$ with $0$.
\State \hspace{2ex} \textbf{for} $i \in \{m-1, m-2, \dots, 0\}$ \textbf{do} \textit{(in order)}
\State \hspace{4ex} \textbf{if} $c_{i+1} = 0$ \textbf{then}
\State \hspace{6ex} Set $\textup{add\_index} = i$.
\State \hspace{4ex} \textbf{else}
\State \hspace{6ex} Set $\textup{add\_index} = i - x$.
\State \hspace{4ex} \textbf{end if}
\State \hspace{4ex} \textbf{if} $c_i = 1$ \textbf{then}
\State \hspace{6ex} $g(\bold{c}) \leftarrow g(\bold{c}) + N(\textup{add\_index}, x)$.
\State \hspace{4ex} \textbf{end if}
\State \hspace{2ex} \textbf{end for}
\State \hspace{2ex} Compute $\bold{b}=\textup{binary}(g(\bold{c})-1)$, which has length $s^{\textup{c}}$. \textit{(decimal integer to binary sequence)}
\State \hspace{2ex} Ignore the next $x$ bridging bits.
\State \textbf{end for}
\State \textbf{Output:} Outgoing stream of binary messages.
\end{algorithmic}
\label{alg_dec}
\end{algorithm}

Since Algorithm~\ref{alg_dec} is a direct consequence of Theorem~\ref{thm_rule}, we refer the reader to Example~\ref{ex_2} for more understanding of how the algorithm works.

On the level of a single message-codeword pair, there exists a single \textbf{for} loop on $m$ distinct values for the variable $i$ in both Algorithm~\ref{alg_enc} (the encoding algorithm) and Algorithm~\ref{alg_dec} (the decoding algorithm). For each value of $i$, at most one major arithmetic operation is performed. Thus, the complexity of both algorithms has $O(m)$ on that level. Moreover, the main operations in Algorithm~\ref{alg_enc} are comparisons/subtractions, while the main operations in Algorithm~\ref{alg_dec} are additions. The largest result of these operations is the maximum value the index $g(\bold{c})$ can take, which is $2^{s^\textup{c}}-1$. Consequently, the size of the used adders, which is $s^\textup{c}$, dictates the complexity of the encoding and decoding algorithms of CA-LOCO codes.

Table~\ref{table_3} links various finite-length rates of CA-LOCO codes $\mathcal{AC}^{\textup{c}}_{m,x}$ for different values of $m$ and $x \in \{1, 2\}$ to the associated size of adders required to achieve these rates, which is a crucial complexity measure. For example, for the case of $x=1$, to achieve a rate $\geq 0.8000$ (resp., $\geq 0.8050$), adders of size $36$ bits (resp., $62$ bits) suffice. Moreover, for the case of $x=2$, to achieve a rate $\geq 0.6667$ (resp., $\geq 0.6800$), adders of size $20$ bits (resp., $45$ bits) suffice. Note that the two cases of $\mathcal{AC}^{\textup{c}}_{357,1}$ and $\mathcal{AC}^{\textup{c}}_{244,2}$ are given in the table only to show how close to capacity CA-LOCO codes can get; that is why the adder size is skipped for both.

Next, we compare A-LOCO codes with other constrained codes used for the same purpose. First, we compare with constrained codes based on finite-state machines (FSMs) and sliding window decoders. FSM-based constrained codes are designed by developing an FSTD that represents infinite sequences satisfying the required constraint. Then, multiple steps are performed to generate the encoding-decoding FSM from the FSTD \cite{siegel_mr, siegel_const}. To construct FSM-based constrained codes with capacity-approaching rates, typically the FSM gets quite complicated, and so are the encoding and decoding procedures.

As a result, we compare with FSM-based $\mathcal{A}_x$-constrained codes that are known to be practical in terms of complexity. A practical FSM-based $\mathcal{A}_1=\{101\}$-constrained code has a rate of $0.8000$ \cite{siegel_const}, while practical CA-LOCO codes that are $\mathcal{A}_1$-constrained achieve rates $\geq 0.8050$ at moderate lengths. Additionally, a practical FSM-based $\mathcal{A}_2=\{101, 1001\}$-constrained code has a rate of $0.6667$, while practical CA-LOCO codes that are $\mathcal{A}_2$-constrained achieve rates $\geq 0.6800$ at moderate lengths. The gain in rate achieved by low-complexity CA-LOCO codes (with adder sizes $\leq 64$ bits) compared with practical FSM-based constrained codes reaches $3\%$, which is a significant rate increase for high rates. Techniques raising the rate with similar or less amounts are highly appreciated in the literature, e.g., raising the rate of FSM-based constrained codes forbidding the patterns in $\{0101, 11101\}$ from $\frac{5}{6}$ to $\frac{6}{7}$, which gives a $2.85\%$ gain \cite{siegel_const}.

A-LOCO codes also have other advantages over FSM-based constrained codes designed for the same purpose. A-LOCO codes are fixed-length codes. Thus, they do not allow errors to propagate from a codeword into another. Additionally, they also enable parallel encoding and decoding in applications where runtime operations speed is critical.

Second, we compare with the binary asymmetric constrained codes in \cite{chee_qlc}. In \cite{chee_qlc}, the encoding and decoding are based on the unrank and rank procedures described in \cite[Algorithm~1]{chee_qlc} and \cite[Algorithm~2]{chee_qlc}. While these codes are also enumerative, and thus, can achieve high rates, their encoding and decoding algorithms are more complex than those of A-LOCO codes, which are based on a simple rule described in Theorem~\ref{thm_rule}. Additionally, the codes in \cite{chee_qlc} only consider the effect of adjacent Flash cells, i.e., can only eliminate the pattern $101$ for SLC Flash devices.

Third, we compare with symmetric LOCO codes used for the same goal. In particular, we compare an $\mathcal{A}_x$-constrained code (A-LOCO code) of length $m$ with the $\mathcal{S}_x$-constrained code (LOCO code) of length $m$. For Flash devices where the goal is only to eliminate the patterns in $\{101, 1001, \allowbreak \dots, 1\bold{0}^x1\}$, the gain in rate achieved by A-LOCO codes over LOCO codes at the same low complexity and achieving nearly the same performance reaches $16\%$ (resp., $25\%$) for $x=1$ (resp., $x=2$). This gain is expected knowing that the capacity of $\mathcal{S}_1$-constrained (resp., $\mathcal{S}_2$-constrained) codes is $0.6942$ (resp., $0.5515$) \cite{ahh_loco}. More details are available in Table~\ref{table_3} and \cite[Table~IV]{ahh_loco}. We note that a similar observation was stated in \cite{siegel_const} in the context of MR systems.

Like LOCO codes \cite{ahh_loco}, A-LOCO codes are reconfigurable, that is, the same hardware can be used to support multiple constraints if the size of the adders is appropriately chosen.

%%%%%%%%%%%%%%%%%%%%%%%%%%%%%%%%%%%%%
\section{Conclusion}\label{sec_conc}

We introduced a new family of asymmetric constrained codes, A-LOCO codes, to improve the performance in Flash memories. Only the detrimental patterns in Flash systems are eliminated in A-LOCO codewords. We derived a recursive formula to compute the cardinality of A-LOCO codes. We presented a simple rule for the mapping-demapping between the lexicographic index and the codeword. This rule allowed practical, low-complexity encoding and decoding algorithms of A-LOCO codes. We illustrated how to optimally bridge and to self-clock A-LOCO codes. We showed that A-LOCO codes are capacity-achieving. The complexity of encoding-decoding A-LOCO codes was studied and comparisons with other constrained codes were presented. These comparisons demonstrated that A-LOCO codes offer a rate-complexity trade-off that is better than other constrained codes used for the same purpose. Non-binary constrained codes for Flash devices with more than two levels and multi-dimensional constrained codes for multi-dimensional storage devices are among the near future research directions. QLC Flash memory evolution is expected to benefit from efficient high rate non-binary asymmetric constrained codes.

%%%%%%%%%%%%%%%%%%%%%%%%%%%%%%%%%%%%%
\section*{Acknowledgment}\label{sec_ack}

This research was supported in part by NSF under grant CCF 1717602.

%%%%%%%%%%%%%%%%%%%%%%%%%%%%%%%%%%%%%
\balance


\begin{thebibliography}{13}

\bibitem{tang_bahl}
D. T. Tang and R. L. Bahl, ``Block codes for a class of constrained noiseless channels,'' \emph{Inf. and Control}, vol. 17, no. 5, pp. 436--461, 1970.

\bibitem{siegel_mr}
P. Siegel, ``Recording codes for digital magnetic storage,'' \emph{IEEE Trans. Magn.}, vol. 21, no. 5, pp. 1344--1349, Sep. 1985.

\bibitem{vasic_prc}
B. Vasic and E. Kurtas, \textit{Coding and Signal Processing for Magnetic Recording Systems.} CRC Press, 2005.

\bibitem{immink_1}
K. A. S. Immink, P. H. Siegel, and J. K. Wolf, ``Codes for digital recorders,'' \emph{IEEE Trans. Inf. Theory}, vol. 44, no. 6, pp. 2260--2299, Oct. 1998.

\bibitem{immink_opt}
K. A. S. Immink, `` Modulation systems for digital audio discs with optical readout,'' in \emph{Proc. IEEE Int. Conf. Acoust., Speech, Signal Process. (ICASSP)}, Atlanta, Georgia, USA, Mar.--Apr. 1981, pp. 587--589.

\bibitem{robert_spec1}
G. D. Forney and A. R. Calderbank, ``Coset codes for partial response channels; or, coset codes with spectral nulls,'' \emph{IEEE Trans. Inf. Theory}, vol. 35, no. 5, pp. 925--943, Sep. 1989.

\bibitem{qin_flash}
M. Qin, E. Yaakobi, and P. H. Siegel, ``Constrained codes that mitigate inter-cell interference in read/write cycles for flash memories,'' \emph{IEEE J. Sel. Areas Commun.}, vol. 32, no. 5, pp. 836--846, Apr. 2014.

\bibitem{kayser_flash}
S. Kayser and P. H. Siegel, ``Constructions for constant-weight ICI-free codes,'' in \emph{Proc. IEEE Int. Symp. Inf. Theory (ISIT)}, Honolulu, HI, USA, Jun.--Jul. 2014, pp. 1431--1435.

\bibitem{veeresh_mlc}
V. Taranalli, H. Uchikawa, and P. H. Siegel, ``Error analysis and inter-cell interference mitigation in multi-level cell flash memories,'' in \emph{Proc. IEEE Int. Conf. Commun. (ICC)}, London, UK, Jun. 2015, pp. 271--276.

\bibitem{chee_qlc}
Y. M. Chee, J. Chrisnata, H. M. Kiah, S. Ling, T. T. Nguyen, and V. K. Vu, ``Capacity-achieving codes that mitigate intercell interference and charge leakage in Flash memories,'' \emph{IEEE Trans. Inf. Theory}, vol. 65, no. 6, pp. 3702--3712, Jun. 2019.

\bibitem{siegel_const}
R. Karabed and P. H. Siegel, ``Coding for higher-order partial-response channels,'' in \emph{Proc. SPIE 2605, Coding and Signal Process. for Inf. Storage}, Philadelphia, PA, USA, Dec. 1995, pp. 115--127.

\bibitem{cover_lex}
T. Cover, ``Enumerative source encoding,'' \emph{IEEE Trans. Inf. Theory}, vol. 19, no. 1, pp. 73--77, Jan. 1973.

\bibitem{immink_lex}
K. A. S. Immink, ``A practical method for approaching the channel capacity of constrained channels,'' \emph{IEEE Trans. Inf. Theory}, vol. 43, no. 5, pp. 1389--1399, Sep. 1997.

\bibitem{braun_lex}
V. Braun and K. A. S. Immink, ``An enumerative coding technique for DC-free runlength-limited sequences,'' \emph{IEEE Trans. Commun.}, vol. 48, no. 12, pp. 2024--2031, Dec. 2000.

\bibitem{ahh_loco}
A. Hareedy and R. Calderbank, ``LOCO codes: lexicographically-ordered constrained codes,''  \emph{IEEE Trans. Inf. Theory}, to be published, doi: 10.1109/TIT.2019.2943244.

\bibitem{ahh_tit2}
A. Hareedy, R. Wu, and L. Dolecek, ``A channel-aware combinatorial approach to design high performance spatially-coupled codes for magnetic recording systems,'' Sep. 2018. [Online]. Available: https://arxiv.org/abs/1804.05504

\bibitem{laroia_const}
R. Laroia, N. Farvardin, and S.A. Tretter, ``On optimal shaping of multidimensional constellations,'' \emph{IEEE Trans. Inf. Theory}, vol. 40, no. 4, pp. 1044--1056, Jul. 1994.

\end{thebibliography}
\end{document}